\documentclass[aps,prl,twocolumn,superscriptaddress]{revtex4-2}

\usepackage{hyperref}
\usepackage{graphicx}
\usepackage{siunitx}
\usepackage[normalem]{ulem} 

\usepackage[pdftex,dvipsnames,usenames]{color}	

\newcommand{\Figref}[1]{Fig.~\ref{fig:#1}}
\usepackage{subcaption}
\def\figref#1{\ref{fig:#1}}
\def\figlab#1{\label{fig:#1}}  

\begin{document}


\title{Filamentary dynamics in fan-like structures in high-altitude lightning discharges}


\author{M.A.A. \surname{Lourens}}
\email[Contact author: ]{mlourens@astron.nl}
\affiliation{Netherlands Institute for Radio Astronomy (ASTRON), Dwingeloo, Netherlands}
\affiliation{Kapteyn Astronomical Institute, University of Groningen, Groningen, Netherlands}

\author{B.M. \surname{Hare}}
\email[Contact author: ]{hare@astron.nl}
\affiliation{Netherlands Institute for Radio Astronomy (ASTRON), Dwingeloo, Netherlands}
\affiliation{Kapteyn Astronomical Institute, University of Groningen, Groningen, Netherlands}

\author{O. \surname{Scholten}}
\affiliation{Netherlands Institute for Radio Astronomy (ASTRON), Dwingeloo, Netherlands}
\affiliation{Kapteyn Astronomical Institute, University of Groningen, Groningen, Netherlands}

\author{S. \surname{Cummer}}
\affiliation{Department of Electrical and Computer Engineering, Duke University, Durham, NC, USA}

\author{J. \surname{Dwyer}}
\affiliation{Department of Physics \& Astronomy and Space Science Center (EOS), University of New Hampshire, Durham, NH, USA}

\author{N. \surname{Liu}}
\affiliation{Department of Physics \& Astronomy and Space Science Center (EOS), University of New Hampshire, Durham, NH, USA}

\author{C. \surname{Sterpka}}
\affiliation{Netherlands Institute for Radio Astronomy (ASTRON), Dwingeloo, Netherlands}
\affiliation{Kapteyn Astronomical Institute, University of Groningen, Groningen, Netherlands}

\author{P. \surname{Turekova}}
\affiliation{Netherlands Institute for Radio Astronomy (ASTRON), Dwingeloo, Netherlands}
\affiliation{Kapteyn Astronomical Institute, University of Groningen, Groningen, Netherlands}

\author{B. \surname{Wu}}
\affiliation{State Key Laboratory of Severe Weather Meteorological Science and Technology \& CMA Key Laboratory of Lightning, Chinese Academy of Meteorological Sciences, Beijing, China}

\author{S. \surname{ter Veen}}
\affiliation{Netherlands Institute for Radio Astronomy (ASTRON), Dwingeloo, Netherlands}


\begin{abstract}
We present LOFAR radio telescope measurements revealing that high-altitude negative leaders (HANLs) exhibit behavior completely different from the current paradigm of negative leader propagation. Our data show that the fan at the tip of a propagating HANL is formed by filaments that behave like smoothly advancing negative leaders. These negative leader filaments appear to have moderate conductivity, remain thin (radius $\sim\SI{1}{\meter}$), grow continuously at $\sim\SI{e5}{\meter\per\second}$ over hundreds of meters, and emit VHF radiation persistently. The observations show that the filaments making up the fan in front of a propagating negative leader are not necessarily streamers, as commonly assumed. Previous work suggesting to observe streamers at the tips of negative leaders should be re-examined in light of these new findings.
\end{abstract}

\maketitle

Lightning is a well-known but poorly understood electrical discharge process. The discharge proceeds by forming hot conducting channels with temperatures exceeding \SI{5\,000}{\kelvin}, called leaders, propagating over distances of many kilometers through the atmosphere and polarizing in the ambient electric field\cite{Dwyer2014, daSilva2019, Wemhoner2026}. The positively charged leader-end generally propagates in a continuous fashion \cite{Berger1967}, whereas the negatively charged leader-end is known to always propagate in a sequence of jumps commonly referred to as steps \cite{Schonland1935, Berger1967}. Studying the physics of in-cloud negative leader propagation can be performed very effectively through the detection of short radio pulses in the VHF frequency regime, that are emitted while the discharge propagates. Leveraging the high-spatial resolution of the LOFAR radio telescope \cite{Haarlem2013} allows investigation of the dynamics of this stepping process. Using a recently developed, even more refined, imaging technique~\cite{Scholten2025}, we observed an unexpected mode of electrical breakdown while investigating the stepping mode of high-altitude negative leaders (HANLs)~\cite{Pu2024, Scholten2021}.

In HANLs these steps show in VHF images as a series of fan-like structures, see \cite{Lyu2016, Pu2024, Scholten2021},  consisting of multiple filaments. These fans appear every few milliseconds separated by up to \SI{300}{\meter}. Long filaments on HANLs have also been captured in two rare long optical exposures of a HANL emerging from a cloud in \cite{Edens2014}. These filaments have always been interpreted as a corona of streamers, a type of self-propagating discharge \cite{Raizer1991, Nijdam2020}, forming at the propagating tip of the leader where the electric field is sufficiently enhanced to produce streamer discharges. However, due to the lack of detailed observations, the true nature of these filaments remained an open question.

In this work, we present the highest resolution image of HANL propagation (see Fig.~\figref{HANL_fan}) where we investigate the filaments inside a fan-structure \footnote{See Supplemental Material [URL will be inserted by publisher] for a 3D animation of the development of the fan-structure presented in this letter.}. These images show unambiguously that these filaments cannot be streamers -as commonly assumed- but are instead leader-like discharges.
The filaments are 100 to 500 meter long negatively-charged structures with an observed diameter of the order of our $\sim\SI{1}{\meter}$ instrumental resolution with no sign of broadening while propagating over such distances through virgin air. They are observed to propagate with a nearly constant speed of $\sim\SI{e5}{\meter\per\second}$. Branching of these filaments is common, with no noticeable change in propagation speed. They are observed through the rather constantly emitted, short duration VHF-pulses from the tip of the filament. The observations show evidence that their tracks are moderately conductive. Later filaments start on the track of previous filaments, but appear to avoid the earlier tracks. 

The observed behavior differs from that of individual streamers and favors a leader-like interpretation, where streamers are self propagating charges that leave no conducting channel behind (since there is no ohmic heating) while leaders do. It is generally accepted that negative leaders step; however, the full connection between charge and stepping behavior is a major open question in lightning research. Our present observations show, in contrast, that filaments are commonly occurring negative leaders that propagate in a non-stepping mode over distances of \SI{500}{\meter} without perceivable broadening, matching the self-focusing characteristic expected from leaders.

The HANL we focus on in this work is part of a lightning event recorded on August 14, 2020 at 14:14:58 UTC by 200 antennas across 36 of the 38 LOFAR stations in the northern Netherlands. The antennas, located mainly around the Exloo core with baselines up to \SI{100}{\kilo\meter}, operate in the $\SI{30}{}-\SI{80}{\mega\hertz}$ VHF band. In this frequency range, lightning propagation leads to the emission of millions of short radio pulses, most originating from the propagating tips of negative leaders. We re-imaged the development of the negative leader using our recently developed adaptive time-resolved interferometric 3-dimensional (ATRI-D) imager \cite{Scholten2025}.

The ATRI-D imager is a beamforming imager that coherently sums the VHF radio signals received by hundreds of antennas. Key is that this beamformer can handle signals from antennas with various viewing angles to the source \cite{Scholten2022, Scholten2025}. ATRI-D has proven to be able to locate weak sources on the body of an aircraft flying through high clouds that were indistinguishable from noise on a single antenna with precision on the order of \SI{10}{\centi\meter} \cite{Scholten2025}. Part of the selected HANL has been studied before with much lower sensitivity and precision in \cite{Scholten2021} using the impulsive imager, which locates pulses using a Kalman filtering inspired time-of-arrival imaging technique \cite{Scholten:2021-init}.

An overview of the imaged section of the HANL is shown in Fig.~\figref{HANL_overview}.
\begin{figure*}
   \begin{subfigure}[t]{0.49\textwidth}
       \includegraphics[width=\linewidth]{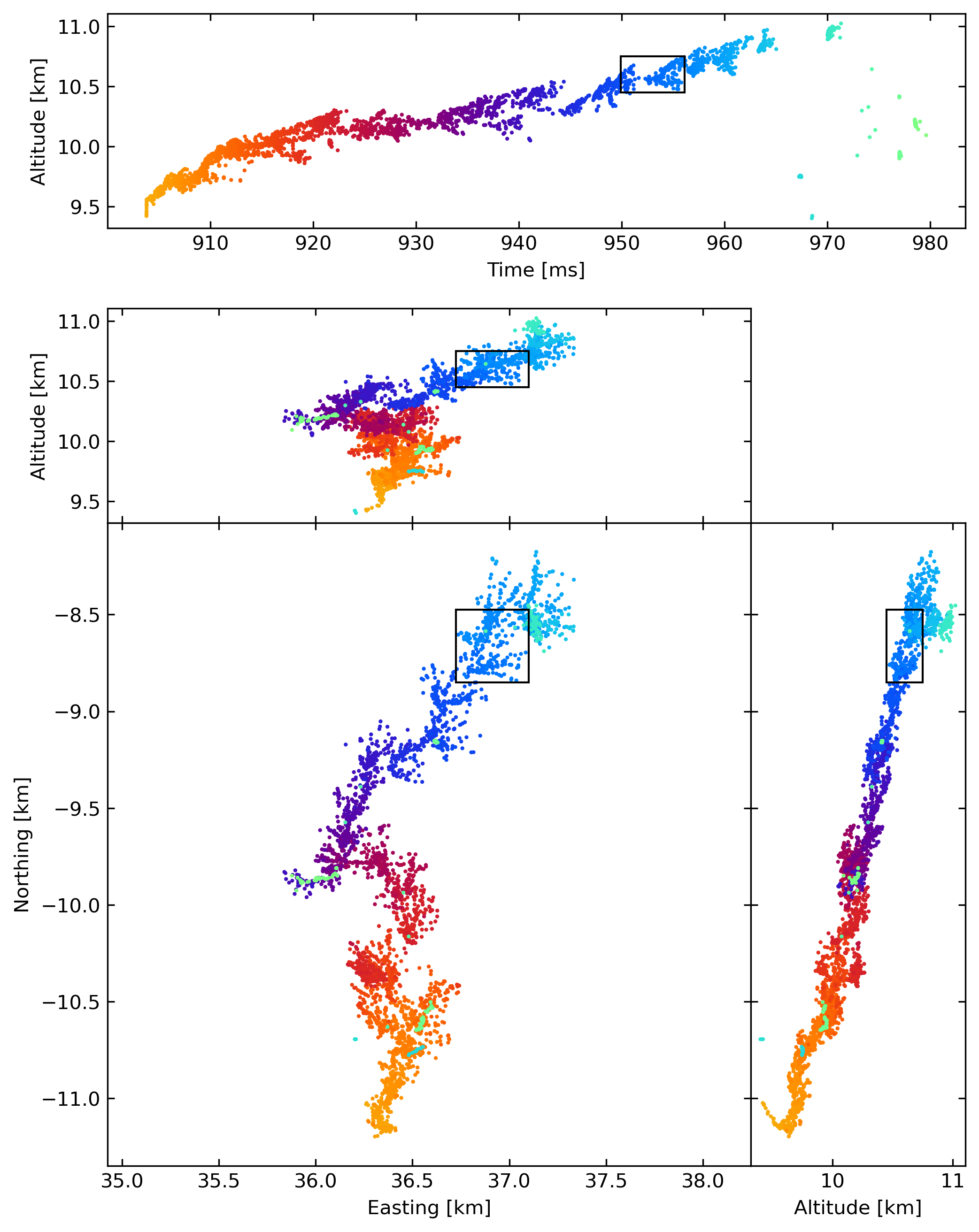}
   	 \caption{}\figlab{HANL_overview}
   \end{subfigure}
   \begin{subfigure}[t]{0.49\textwidth}
        \includegraphics[width=\linewidth]{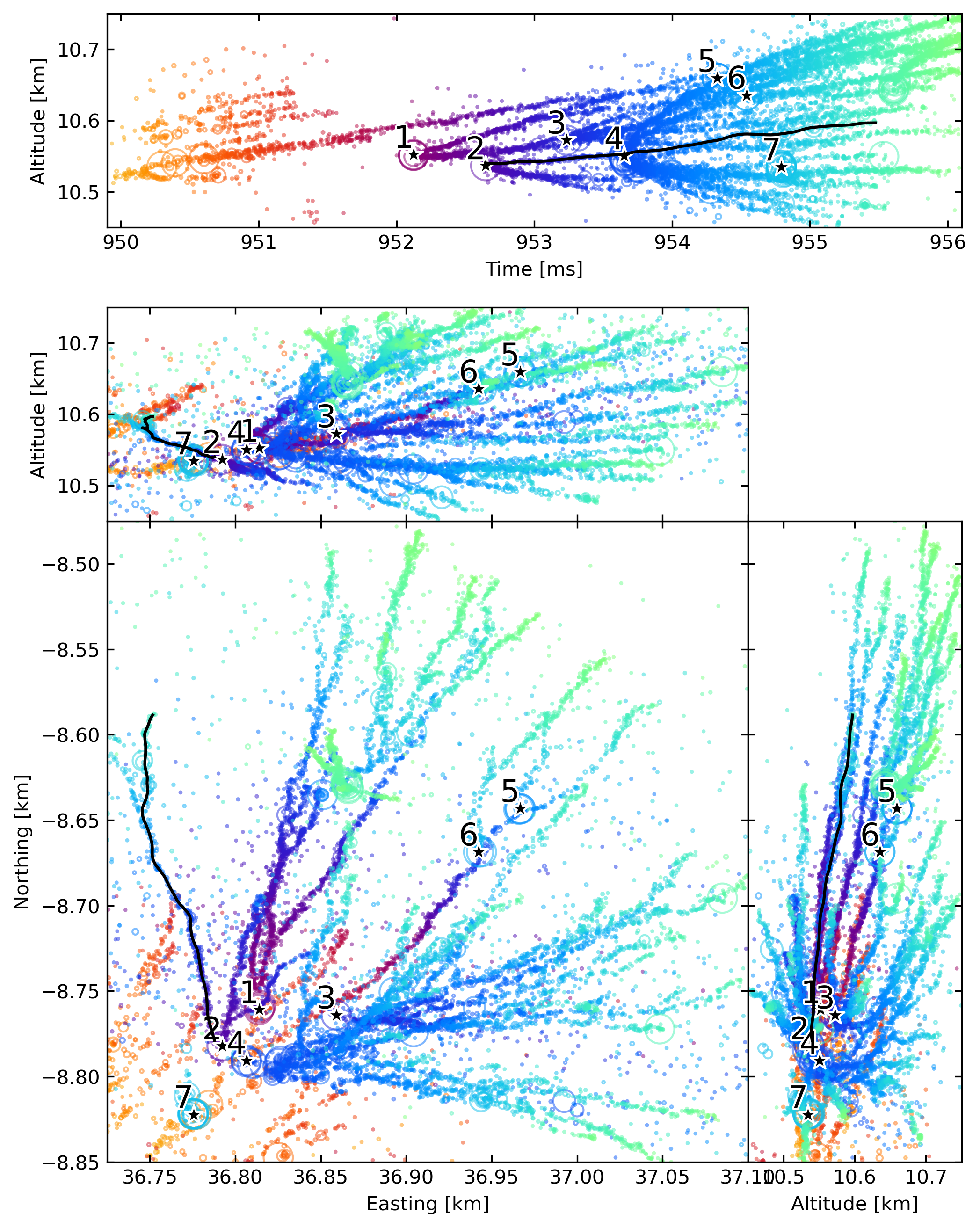}
   	 \caption{}	 \figlab{HANL_fan}
   \end{subfigure}
    \caption{Section of high-altitude negative leader (HANL) (\figref{HANL_overview}) and enhanced image of individual corona fan (\figref{HANL_fan}). Points indicate location of VHF sources. Top-panels show time-evolution of discharge in altitude and bottom-panels show planar views of the development (color indicates time). Fig.~\figref{HANL_overview} shows that HANLs propagate as a series of fans. The indicated region in Fig.~\figref{HANL_overview} is re-imaged using ATRI-D and shown in Fig.~\figref{HANL_fan}. The scale of points reflects spectral energy density of the pulse at \SI{60}{\mega\hertz}. The most intense bursts of VHF emission mark the start of new networks of filaments. A selection of bursts occurring along a filament are marked by stars and numbered in order of occurrence. Fig.~\figref{HANL_fan} shows that fans consist of an intricate network of narrow filaments starting at various times and locations. The time averaged position of an individual filament starting with burst 2 obtained using sliding Gaussian kernel with $\sigma_t = \SI{50}{\micro\second}$ is indicated by a black line. The scatter of VHF sources around the average track is analyzed in Fig.~\ref{fig:filament_properties}.}
\end{figure*}
This section is part of a larger negative leader structure that initiated at \SI{861.4}{\milli\second} at $\left(N, E, h\right) = \left(-10.6, 36.4, 9.0\right)$ \SI{}{\kilo\meter} (outside the volume and before the data shown in Fig.~\figref{HANL_overview}). All coordinates used are relative to LOFAR's core near Exloo, Netherlands. The panels in Fig.~\figref{HANL_overview} show the time evolution in 3D of the discharge (time is colored) consisting of multiple fan-like structures (hereafter referred to as ``fans'') that propagate the negative leader forward in steps. The fans have different sizes, up to \SI{300}{\meter} long, and grow in milliseconds. The fan-like structures in Fig.~\figref{HANL_overview} appear very similar to the results presented in \cite{Edens2014, Pu2024,Yang2025, Yang2025a}. However, there are some exceptions. The results in \cite{Pu2024, Yang2025, Yang2025a} show that the strongest VHF emission within the fan is associated with a fast propagating discharge extending at $\sim\SI{e7}{\meter\per\second}$ over a distance of $\sim\SI{100}{\meter}$ before initiating a diffuse cloud of emission from its end. The fan presented in this work develops without fast propagation and the most intense VHF emission is localized, likely emitted from a region $\sim\SI{1}{\meter}$ in size.

An ATRI-D image of the boxed region in Fig.~\figref{HANL_overview} is shown in Fig.~\figref{HANL_fan}. Fig.~\figref{HANL_fan} reveals that on a smaller scale, the fans do not grow as a single structure. Instead, they consist of an intricate network of thin filaments ($<\SI{1.1}{\meter}$ wide) that grow with average speeds of $\sim\SI{e5}{\meter\per\second}$ at different times from various locations. A propagation speed of \SI{0.8e5}{\meter\per\second} can easily be deduced from this figure as the filaments cover a distance of about \SI{250}{\meter} in \SI{3}{\milli\second}. The width in each dimension of a single filament indicated in Fig.~\figref{HANL_fan} is summarized in Fig.~\figref{filament_properties}. The three panels of Fig.~\figref{filament_properties} show distributions of the deviations in easting, northing and altitude of source locations from the black track indicated in Fig.~\figref{HANL_fan}. The scatter in these three panels shows the progression of the spread in sources around the track as a function of time. The broadening at \SI{954.5}{\milli\second} in the easting and altitude panels is due to branching of the filament.
\begin{figure}
    \includegraphics[width=\linewidth]{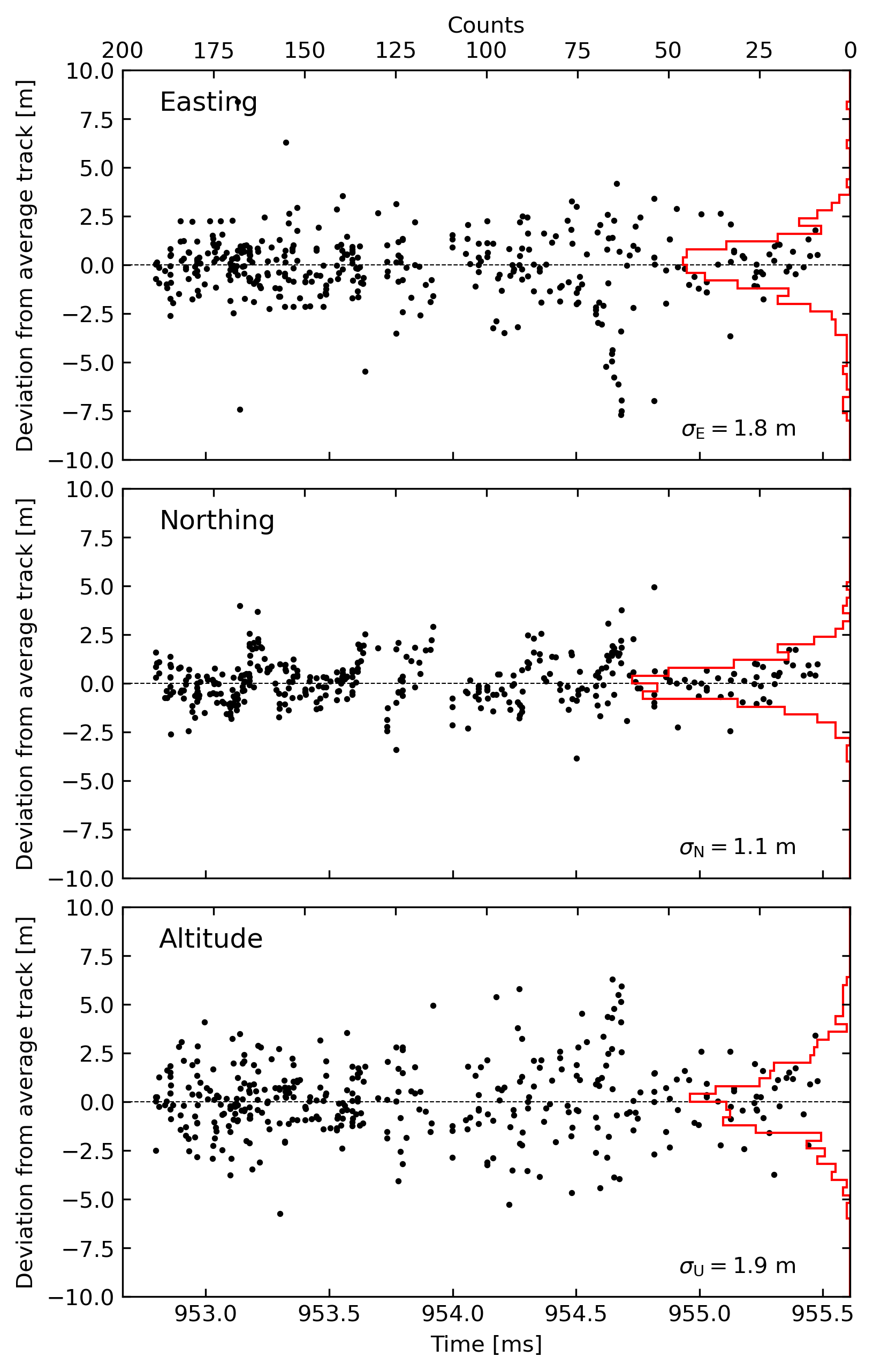}
    \caption{Deviations of located VHF sources from the average track indicated in Fig.~\figref{HANL_fan} as histograms and scatter plots in time. The three histograms show the deviations of VHF source locations from the average location of the filament in easting, northing, and altitude, which are indicated using the vertical axis. A scatter showing the time evolution of these deviations is plotted on top of the histograms with time indicated on the top abscissa. The broadening of the easting and altitude scatter at \SI{954.5}{\milli\second} is due to the filament branching. The RMS deviation from the average track in each dimension is indicated in the bottom right corner of each panel.\figlab{filament_properties}}
\end{figure}
The width of the filament derived from Fig.~\ref{fig:filament_properties} is only an upper limit since the image resolution is very difficult to estimate. It is expected to be much worse that the \SI{10}{\centi\meter} advocated in \cite{Scholten2025} because of the huge number of (almost) simultaneous sources creating an enhanced noise-background for the source of interest. Based on the large number of stray background sources, the observed spread could easily be dominated by the imaging resolution.

Fig.~\figref{HANL_fan} shows that the filaments propagate away from and bend around any previously formed structures. The filaments do not merge or pass twice through the same volume. The time coloring in the northing versus altitude panel in Fig.~\figref{HANL_fan} shows that filaments growing from location 4 bend around previously formed filaments.

The filaments in a fan do not all originate from one position and time, as Fig.~\figref{HANL_overview} suggests. Instead, Fig.~\figref{HANL_fan} shows that filaments in one fan form multiple separate networks, where each filament network has a different origin marked by a strong VHF pulse (burst) that is more energetic (as indicated by the size of the markers) than other pulses emitted around the same time from that region. The seven bursts indicated in Fig.~\figref{HANL_fan} all occur along filaments that grew from a preceding fan and are numbered in order of occurrence and marked by black stars. Bursts 1, 2, and 7 occur along the same filament, while bursts 3, 4, 5, and 6 occur along a side-branch of this filament. These starting points are also indicated in the animation provided in Supplementary Material~\cite{Note1}. Interestingly, the time coloring shows that bursts and the resulting filament growth do not begin at the tips of filaments formed previously.

The observations suggest that filaments build up charge at distinct positions along their length on a timescale ranging up to milliseconds. This is evidenced by later bursts and subsequent filament growth from the body of the preceding filament. In addition, the irregular occurrence of bursts at various locations along the preceding filament suggests that even after the first burst, charging continues at the later burst locations (see Fig.~\figref{HANL_fan}). This leads us to conclude that there is a charging current flowing through the filament. However, the conductivity must be rather limited; otherwise, charge would not collect at different spots on the filament.

We rule out the possibility that the observed filaments, such as the ones shown in \Figref{HANL_fan}, are individual negative streamers. The first argument is that we observe VHF emission from the propagating tip of a filament. A single streamer propagating at constant speed cannot emit copious VHF pulses \cite{Shi2019, Liu2019, Malla2024}. This implies that multiple streamers need to be present at the propagating tips of the filaments in order to have the observed VHF emission. Another indication is the small number of filaments we observe; streamers are expected to be produced at a frequency of $\sim\SI{e9}{\per\second}$ in the high electric field region around the leader tip \cite{Silva2013}, which is obviously significantly more than the number of filaments we observe. Third, the growth of a streamer depends on the imbalance between ionization and electron attachment, which depends on electric field strength \cite{Nijdam2020}. For the streamer to have a nearly constant propagation velocity would require a nearly constant electric field strength close to the field for steady propagation, which, for negative streamers, lies somewhere between $\sim\SI{10}{\kilo\volt\per\centi\meter}$ and the conventional breakdown electric field of $\sim\SI{28.7}{\kilo\volt\per\centi\meter}$ at standard temperature and pressure, depending on the streamer's properties \cite{Qin2014, Guo2022}. Any deviation from this field strength causes the streamer to quickly accelerate or decelerate \cite{Guo2022, Liu2004, Kosar2012}, which we do not observe. Lastly, the rate of electron attachment dominates over the rate of ionization behind the tip resulting in the loss of conductivity within tens of centimeters behind the streamer head \cite{Guo2022}. We thus conclude that an individual filament cannot be a single streamer.

Similarly, we can also rule out the possibility that individual filaments are solely trees of branching streamers. Simulations show that, due to the charge present at the streamer tips, such a system should exhibit a large opening angle \cite{Luque2014, Wang2023, Teunissen2025}. In contrast, our observations show no measurable broadening of the filament as it propagates over distances of $\sim\SI{500}{\meter}$.

Therefore, we conclude that while VHF emission implies there must be significant streamer activity at the tip of every filament, there must also be some kind of self-focusing mechanism that keeps these filaments confined to a small diameter of less than a meter since a bundle of streamers will diverge~\cite{Luque2008a, Luque2014, Silva2013, Carlson2024}. Additionally, the observation of bursts along earlier formed filament segments are evidence of charge accumulation as the filament continued to grow. This implies that filaments must be carrying current and therefore must be somewhat conductive. 
This leaves the intriguing question about the relation between filaments and stepping negative leaders where each fan structure has been interpreted as a step\cite{Edens2014}. In this work we have observed that filaments may propagate from one fan into the following without any change of speed, and that the bursts that start filament trees tend to occur randomly in location and time. 
The result is that we do not see how the commonly accepted interpretation of stepping being driven by a space-leader \cite{Renardieres1974} applies to HANLs.
It could be that space-leaders play a role in heating filaments to higher temperatures, a process that cannot be observed in VHF. 
Instead we have shown that the fan structures are built from multiple, often branching, filaments that may start from different locations along an existing filament. We still do not understand why the total of all filaments make repeated fan-like structures. When a series of fans are imaged with a resolution much worse than presented in this work it gives the impression of a discrete stepping process. There is still the additional question of how negative leaders at lower altitudes (below 7~km) relate to HANLs. It is possible that they propagate in the same fashion as HANLs but simply scaled down due to higher pressure. However, it is also possible that HANL fans are due to environmental factors (e.g., charge pockets) and thus are fundamentally different from negative leader stepping commonly observed near ground (e.g. \cite{Saba2022}).

This view is a critical change in how we believe lightning plasma works. It has always been taken for granted that negative leaders, due to their repelling of free electrons, behave intrinsically different from positive leaders, stepping vs.\ continuous propagation mode. There is still a marked difference between the two. Filaments and hot negative leaders produce copious VHF-emission at their tips, while positive leaders are very VHF-quiet \cite{Scholten2023a}. Negative filaments and non-stepping positive leaders share the self-focusing characteristic. One expects the discharge to increase in diameter during propagation, as like charges repel. However, for leaders, heating will be largest near the center of the tip (due to streamer activity). Once the temperature exceeds a critical value, ohmic heating from the advancing charge through the more conductive part of the tip will also start heating the channel. The result is a type of heating-feedback, thus concentrating heat (and therefore conductivity) to a limited width. \cite{Lalande:2002, Silva2013, Carlson2024}

Negative leaders have also been imaged in high resolution at optical wavelengths. The similarities between the structures observed in this study and the dim optical emission in front of a leader seen previously (e.g., \cite{Edens2014, Saba2022}), which is commonly attributed to streamer coronas, show that a fresh look is required to reconcile VHF and optical observations. It is entirely possible that these dimmer channels are leaders. Other work has shown that lightning leaders have a wide range of temperatures, ranging from \SI{2\,000}{\kelvin} (streamer-to-leader transition) \cite{Silva2013} to well over \SI{15\,000}{\kelvin} \cite{Dwyer2014, daSilva2019, Wemhoner2026}. According to \cite{Naghizadeh-Kashani2002, daSilva2019}, the net emission coefficient for an optically thin plasma can change by 10 orders of magnitude in this temperature range.

Although LOFAR is capable of reconstructing 3D emitting dipole orientations by measuring polarization, interpreting these results is complex, so they are omitted from this study. Nonetheless, a preliminary analysis of a single filament indicates that the emitting dipoles appear randomly oriented. We plan to perform a more detailed study of filaments to see how similar their behavior is in different environments and check whether lower altitude negative leaders show similar structures. In addition, simultaneous high-speed optical observations are required to build better understanding of leader development and relate our observations to other work. In \cite{Montanya2015}, a smoothly propagating negative channel is observed in optical high-speed video, which seems to propagate with the same speed we observe. However, currently there is no way to say whether these are, in-fact, the filaments we observe. Lastly, advancements in modeling are needed to explain the formation and growth of filaments.

\begin{acknowledgments}
ML, BMH, CS, and PT are supported by ERC [Grant Agreement No. 101041097]; NL acknowledges the AFOSR Award FA9550-24-1-0124 from the University of New Hampshire; BW is supported by the National Natural Science Foundation of China (Grant 42275090).\\
This paper is based on data obtained with the LOFAR telescope (LOFAR-ERIC) under project code LT14\_001. LOFAR \cite{Haarlem2013} is the Low Frequency Array designed and constructed by ASTRON. It has observing, data processing, and data storage facilities in several countries, that are owned by various parties (each with their own funding sources), and that are collectively operated by the LOFAR European Research Infrastructure Consortium (LOFAR-ERIC) under a joint scientific policy. The LOFAR-ERIC resources have benefited from the following recent major funding sources: CNRS-INSU, Observatoire de Paris and Université d'Orléans, France; Istituto Nazionale di Astrofisica (INAF), Italy; BMBF, MIWF-NRW, MPG, Germany; Science Foundation Ireland (SFI), Department of Business, Enterprise and Innovation (DBEI), Ireland; NWO, The Netherlands; The Science and Technology Facilities Council, UK; Ministry of Science and Higher Education, Poland.
\end{acknowledgments}

\bibliography{bibliography.bib}

\end{document}